\documentclass[aps,12pt,nofootinbib,floatfix,longbibliography]{revtex4-2}

\usepackage{amsmath,amssymb}
\usepackage{graphicx}
\usepackage{bm}
\usepackage[hidelinks]{hyperref}

\begin{document}

\title{The Quantum Noise Fraction and the Addressable Fraction\\in High-Frequency Gravitational Wave Detection}

\author{Sergio Gaudio}
\email{gaudios@laccd.edu}
\affiliation{Department of Physics, Los Angeles City College, Los Angeles, CA 90029, USA}

\begin{abstract}
The quantum noise fraction $\beta$, the share of a detector's noise power that is quantum in origin, sets a ceiling $\mathcal{E}_{\mathrm{max}} = 1/\sqrt{1-\beta}$ on the sensitivity improvement that removing quantum-origin noise can yield. In the kHz--GHz band the readout degree of freedom is a bosonic mode and $\beta = 1/(2\bar{n}_{\mathrm{th}}+1)$, with $\bar{n}_{\mathrm{th}}$ the Bose--Einstein occupation. Quantum noise then dominates only on the cold side of the \emph{thermal frontier} $k_BT\ln 3 = \hbar\omega$ (above 229~MHz at 10~mK), which depends only on the bosonic character of the mode and governs mechanical resonators and electromagnetic cavities alike; a real readout, whose added noise we count as quantum in origin only up to the minimum that quantum mechanics imposes on a linear amplifier, turns it into a joint condition on the occupation and on the excess noise of the readout chain. On a multimode acoustic antenna the frontier can be observed within one device, since the logarithmic slope of $\sqrt{S_hQ}$ along the odd-overtone comb increases by one half across it whatever the dependence of the quality factor on overtone number, which displaces the slopes of $\sqrt{S_h}$ themselves. On this antenna, crossing the frontier costs more in classical sensitivity than quantum enhancement returns. For a GHz mode of an existing bulk acoustic wave device at 10~mK, $\beta = 0.984$ and $\mathcal{E}_{\mathrm{max}} = 7.8$. That ceiling cannot be reached with a linear readout, because with intrinsic damping the zero-point term is the force noise of the bath, fixed by the fluctuation--dissipation theorem and unaffected by any operation on the state of the mode. The \emph{addressable fraction} $\beta_a$ excludes it: at the standard quantum limit the same mode has $\beta = 0.992$ but $\beta_a = 0.496$, an available improvement of 1.41 against a ceiling of 11.0, and less through the readout chains demonstrated at GHz frequencies if their excess noise is technical. An array of $10^4$ such devices cross-correlated for one year reaches $3.6\times 10^{-24}\,\mathrm{Hz}^{-1/2}$ for an integration bandwidth equal to the linewidth; even granting $\mathcal{E}_{\mathrm{max}} = 7.8$ in full, a factor $1.0\times 10^{10}$ above the big-bang-nucleosynthesis bound remains, and it is classical.
\end{abstract}

\maketitle

\noindent{\small This is the Accepted Manuscript version of an article accepted for publication in \emph{Classical and Quantum Gravity}. IOP Publishing Ltd is not responsible for any errors or omissions in this version of the manuscript or any version derived from it. The Version of Record is available online at \href{https://doi.org/10.1088/1361-6382/aeb348}{https://doi.org/10.1088/1361-6382/aeb348}. This Accepted Manuscript is available for reuse under a CC BY-NC-ND licence after the 12 month embargo period provided that all the terms and conditions of the licence are adhered to.}

\section{Introduction}
\label{sec:intro}

Gravitational wave detection above 10~kHz is an active experimental field~\cite{Aggarwal2025}. Proposed detector concepts include electromagnetic cavities in static magnetic fields~\cite{Berlin2022}, superconducting radio-frequency cavities read out as mechanical bars~\cite{Berlin2023}, lumped-element axion haloscopes in static magnetic fields~\cite{Domcke2022}, magnets read out as resonant masses~\cite{DomckeEllisRodd2025}, bulk acoustic wave (BAW) resonators~\cite{Goryachev2014,Goryachev2021}, levitated sensors~\cite{Arvanitaki2013,Aggarwal2022}, optomechanical membranes~\cite{Membrane2025}, haloscope geometries optimized for gravitational waves~\cite{Domcke2024}, and atomic sensors coupled to resonant cavities~\cite{CaiVisinelliYan2025}. Despite rapid experimental progress, current sensitivities remain several orders of magnitude short of realistic astrophysical or cosmological sources~\cite{Aggarwal2025,GrAHal2025}.

Can quantum technologies close this gap? Squeezed vacuum has improved LIGO's sensitivity above 50~Hz by up to 3~dB, raising the expected detection rate by 40--50\%~\cite{Tse2019}, squeezing has since reduced the quantum noise below the standard quantum limit~\cite{Jia2024}, and quantum-limited readout is standard in circuit quantum electrodynamics. What is missing is a diagnostic for \emph{where} in the high-frequency landscape quantum enhancement is effective and where classical noise sets the limit.

A companion paper~\cite{Gaudio2026CQG} established such a diagnostic, in which the maximum enhancement obtainable by removing quantum-origin noise is fixed, under assumptions stated in Sec.~\ref{sec:beta}, by a noise fraction $\beta$ through $\mathcal{E}_{\mathrm{max}} = 1/\sqrt{1-\beta}$. The companion reads $\beta$ as the share of the noise power that quantum techniques can address. For the shot-noise-limited interferometers it treats, that share is the share quantum in origin; for LISA it also includes classical contributions that quantum sensors can reduce, such as oscillator jitter and magnetic forces on the test masses. In the present band the two readings separate, and the fourth result below is their separation. That work classified detectors by the degree of freedom the wave couples to: an internal one, whether atomic state or elastic mode (Mechanism~A); the center-of-mass motion, through Doppler and time-dilation effects (Mechanism~B); or the propagation of light (Mechanism~C). It evaluated $\beta$ for interferometric and atom-interferometric detectors, where it is set by the share of shot or projection noise in the total budget.

The present paper evaluates the same bound in the kHz--GHz band, where the diagnostic becomes more precise. There the degree of freedom read out is a single bosonic mode (a phonon in a resonant mass, a photon in a cavity) and $\beta$ acquires a closed form in the thermal occupation alone, independent of the mass, geometry and quality factor of the device. The consequences are worked out on one platform of the list above, the bulk acoustic wave resonator; the reasons for that choice are given at the start of Sec.~\ref{sec:BAW}.

This band also brings into play a channel the companion deliberately set aside. Its three mechanisms exhaust the ways a wave couples \emph{directly} to a quantum system; indirect mechanisms, in which an external field first converts the wave into another quantity, lie outside its scope, its named example being conversion into photons in a static magnetic field~\cite{Gaudio2026CQG}. A wave traversing a strong static field drives a signal field at the wave frequency, the inverse Gertsenshtein process~\cite{Gertsenshtein1962}, developed for gravitational wave searches in Refs.~\cite{Berlin2022,Domcke2022,Domcke2024,Ratzinger2024}: in the proper detector frame the wave enters as an effective current density $j \sim B_0\,\omega\,h$ for a cavity of size comparable to the wavelength, with $B_0$ the static field, resonantly exciting a cavity mode~\cite{Berlin2022}. We label this Mechanism~D, coupling through conversion in a static background field. It differs in kind from A, B and C, in which the observable responds to the wave itself: in D no material degree of freedom is displaced, the coupling is not tidal, and only the electromagnetic signal produced by the conversion reaches the readout. We do not derive the conversion, for which we rely on Ref.~\cite{Berlin2022}, which corrects and builds upon earlier treatments by accounting for the gauge dependence of the relevant quantities in a preferred laboratory frame, the proper detector frame, and by resumming short-wavelength effects; a frame-independent formulation is given in Ref.~\cite{Ratzinger2024}. What we add is its place in the $\beta$ framework, where it joins Mechanism~A across the direct/indirect divide (Sec.~\ref{sec:frontier}).

Four results follow. First, the closed form places a definite boundary in the frequency--temperature plane, the \emph{thermal frontier} $k_BT\ln 3 = \hbar\omega$, on whose cold side quantum noise is the larger share of the budget. The boundary marks a continuous crossover, and Sec.~\ref{sec:beta} explains why the equal-share point is the natural place for it. Since it depends only on the bosonic character of the readout mode, it governs Mechanisms~A and~D alike. Second, in the device of our case study the frontier can be observed directly. A BAW plate is a multimode acoustic antenna whose odd overtones form a comb spanning up to three decades on one crystal, and the comb changes slope across the frontier; the part of that change that survives the frequency dependence of the quality factor is an invariant equal to one half. Third, the closed form extends naturally to the noise added by the readout chain, once the minimum noise that quantum mechanics imposes on a linear amplifier (Caves' bound, Sec.~\ref{sec:beta}) is used to separate its quantum part, turning the frontier into a joint condition, usable as a design requirement, on temperature and on the excess noise of the readout.

The fourth result is negative and limits what the framework can claim. The ceiling $\mathcal{E}_{\mathrm{max}}$ bounds the gain from removing quantum-origin noise, but in a resonator with intrinsic damping most of that noise cannot be removed. The zero-point term is the force noise of the bath, fixed by the fluctuation--dissipation theorem, and not a property of the state of the mode on which squeezing could act. Separating the removable part defines an \emph{addressable fraction} $\beta_a$, smaller than $\beta$ by the bath's zero-point contribution, and it is $\beta_a$ that bounds what a quantum technique can deliver in such a device.

Taken together, these four results strengthen, for the resonant-mass detectors treated here, the conclusion that the sensitivity deficit at high frequencies is classical. For the acoustic antenna of our case study, crossing the thermal frontier costs more in classical sensitivity than it returns in quantum enhancement, and we make the residual requirement explicit.

\section{The quantum noise fraction}
\label{sec:beta}

For a detector whose total strain noise power spectral density $S_h$ receives contributions from $N$ independent sources, a quantum sensor that reduces source $i$ by a factor $\eta_i$ yields the enhancement~\cite{Gaudio2026CQG}
\begin{equation}
\mathcal{E} = \frac{1}{\sqrt{1 - \sum_i \beta_i(1 - 1/\eta_i^2)}},
\label{eq:enhancement}
\end{equation}
where $\beta_i = S_h^{(i)}/S_h^{\mathrm{total}}$. Even for arbitrarily capable sensors ($\eta_i \to \infty$) acting on every quantum-origin contribution, the enhancement saturates at $\mathcal{E}_{\mathrm{max}} = 1/\sqrt{1-\beta}$ with $\beta = \sum_i \beta_i^{\mathrm{(quantum)}}$. This bound and its consequences for interferometric and atomic detectors are established in Ref.~\cite{Gaudio2026CQG}. Because the bound is used below as a ceiling, we state the assumptions on which it rests. The quantum-origin and classical contributions are taken to be statistically independent and stationary, so that their powers add; correlations within the quantum part, such as those between shot noise and radiation pressure in an interferometer, are allowed, the part being counted as one term. For the bath of a bosonic mode the independence is exact. A thermal state is a Gaussian mixture of coherent states, the vacuum displaced by a random complex amplitude $\alpha$ with probability density $\exp(-|\alpha|^2/\bar{n}_{\mathrm{th}})/\pi\bar{n}_{\mathrm{th}}$~\cite{Glauber1963}, so each quadrature carries the vacuum noise plus an independent classical Gaussian noise of variance $\bar{n}_{\mathrm{th}}$, in units in which the vacuum variance is $1/2$, and the covariance matrix is $2\bar{n}_{\mathrm{th}}+1$ times that of the vacuum~\cite{Weedbrook2012}. The bracket $(2\bar{n}_{\mathrm{th}}+1)$ of Eq.~(\ref{eq:beta_mech}) below is therefore a sum of powers with no cross term. The bound further assumes that each contribution can be reduced without changing the others, a restriction on the operations considered rather than a property of the noise: squeezing the field that enters through a port admitting thermal noise scales the variance $(2\bar{n}+1)/2$ of the squeezed quadrature as a whole, thermal and zero-point parts alike, possibly below the vacuum level, and its gain is not bounded by $\mathcal{E}_{\mathrm{max}}$. No such port exists for a mode whose damping is intrinsic (Sec.~\ref{sec:ceiling}). The technique is taken to act on the noise and not on the signal, so the detector's response to strain, through which every contribution is referred to $S_h$, is held fixed; a change of the transducer changes the budget itself and has to be assessed through a new set of $\beta_i$. The ceiling is reached when every quantum-origin contribution is removed and every classical one is left as it is. Finally, the budget is that of a linear measurement, described by additive spectral densities at the signal frequency. A measurement that is not linear in the field, such as the counting of quanta, lies outside this description (Sec.~\ref{sec:proposal}). Within these assumptions $\mathcal{E}_{\mathrm{max}}$ bounds what the removal of quantum-origin noise can yield. It does not predict what a particular technique will deliver, and Sec.~\ref{sec:ceiling} examines which part of that noise can be removed at all.

What distinguishes the high-frequency regime is that $\beta$ becomes calculable in closed form. For a resonant detector the signal is registered in a single bosonic mode of angular frequency $\omega_0$ in contact with a bath at temperature $T$, and the fluctuation--dissipation theorem fixes the ratio of thermally driven to zero-point contributions with no reference to any other property of the device:
\begin{equation}
\beta = \frac{S_h^{\mathrm{zpf}}}{S_h^{\mathrm{th}} + S_h^{\mathrm{zpf}}} = \frac{1}{2\bar{n}_{\mathrm{th}} + 1} = \tanh u,
\qquad
\bar{n}_{\mathrm{th}} = \frac{1}{e^{\hbar\omega_0/k_BT} - 1},
\qquad
u \equiv \frac{\hbar\omega_0}{2k_BT}.
\label{eq:beta_mech}
\end{equation}
Here $S_h^{\mathrm{zpf}}$ and $S_h^{\mathrm{th}}$ are the strain-noise contributions of the zero-point and thermally driven parts of the fluctuation--dissipation force noise, that is, the terms $1$ and $2\bar{n}_{\mathrm{th}}$ of the bracket $(2\bar{n}_{\mathrm{th}}+1)$, written out in Eq.~(\ref{eq:Sh}) below. The last form follows from $2\bar{n}_{\mathrm{th}}+1 = \coth u$, the fluctuation--dissipation bracket; $u$, rather than $\hbar\omega_0/k_BT$, is therefore the natural dimensionless variable of the problem, and $\beta$ depends on frequency and temperature through it alone. This expression is exact and involves neither the mass, nor the geometry, nor the quality factor, nor the transduction scheme. The fraction $\beta$ is the quantum-origin share of the noise power, which is what Eq.~(\ref{eq:enhancement}) requires; at $\bar{n}_{\mathrm{th}} \ll 1$ the remaining share is the thermal occupation of the same bosonic mode, not classical in any ordinary sense. Whether the quantum share is in practice \emph{addressable} is a further question, which does not arise for the interferometers of the companion, where the quantum share is readout noise and addressable by construction. Equation~(\ref{eq:enhancement}) is smooth in $\beta$ and singles out no value of it: the passage from thermally to quantum dominated noise is a continuous crossover rather than a transition, and any boundary placed on it is a convention. We place it at $\beta = 1/2$, which in Eq.~(\ref{eq:beta_mech}) corresponds to $2\bar{n}_{\mathrm{th}} = 1$. There the zero-point and thermally driven parts of the fluctuation--dissipation force noise contribute equally, the mode is in its ground state with probability $2/3$, and removing all quantum-origin noise would halve the noise power, giving $\mathcal{E}_{\mathrm{max}} = \sqrt{2}$. Among the possible conventions this one has a physical basis. What Eq.~(\ref{eq:enhancement}) does single out is the \emph{variable}, and a threshold stated on $\beta$, rather than on the occupation alone, has the same meaning for every detector whatever the physical origin of its noise. The other natural markers, defined on the state or the energy of the mode and compared in Sec.~\ref{sec:frontier}, lie where the two shares are unequal, by amounts that have no particular meaning for the bound. No conclusion of this paper depends on the choice. Every entry of Table~\ref{tab:landscape} is classified in the same way for any threshold between 0.14, the largest $\beta$ below one half in the table (ADMX), and 0.85, the smallest above it (LIGO), and the case study of Sec.~\ref{sec:proposal}, at $\beta = 0.984$, is quantum dominated for any threshold below that value.

Equation~(\ref{eq:beta_mech}) describes the mode in isolation; a real instrument adds the noise of its readout chain. Writing it, referred to the mode as is standard for a linear measurement chain~\cite{Clerk2010} and evaluated at the mode frequency, as $S_h^{\mathrm{add}} = \nu\,S_h^{\mathrm{zpf}}$, with $\nu$ the added noise in zero-point units, the total becomes
\begin{equation}
S_h^{\mathrm{total}} = (2\bar{n}_{\mathrm{th}} + 1 + \nu)\,S_h^{\mathrm{zpf}} .
\label{eq:total_readout}
\end{equation}
An acoustic mode is read through a transducer and an amplifier chain, a cavity mode through a coupling port and a receiver, and each stage contributes to $\nu$. How much of it is quantum in origin depends on the class of measurement, and this requires a definition of the standard quantum limit (SQL) in the present context. Caves showed that a phase-insensitive linear amplifier, in the limit of high gain, must add noise which, referred to its input, is at least half a quantum, the zero-point level~\cite{Caves1982,Clerk2010}; we refer to this lower limit as Caves' bound. A phase-sensitive amplifier need not add noise, at the price of amplifying only one quadrature while de-amplifying the other~\cite{Caves1982}, so that for a signal of random phase the power in the conjugate quadrature is lost. The units of Eq.~(\ref{eq:total_readout}) are symmetrized (the bracket $(2\bar{n}_{\mathrm{th}}+1)$ of Eq.~(\ref{eq:beta_mech}) is $2(\bar{n}_{\mathrm{th}}+\tfrac12)$, so the bath's zero-point half-quantum already appears as unity), and an added half-quantum enters as $2\times\tfrac12$; the minimum added noise is therefore
\begin{equation}
\nu_{\mathrm{SQL}} = 1 \quad \text{(phase-insensitive)}, \qquad \nu_{\mathrm{SQL}} = 0 \quad \text{(phase-sensitive)},
\label{eq:nu_sql}
\end{equation}
which is Caves' bound in the symmetrized units used throughout. Added noise at that limit is itself quantum in origin and is addressable, by squeezed-state injection or back-action evasion, so it belongs in the numerator of $\beta$; the excess above the limit we book as classical. That bookkeeping is a convention, and a conservative one: part of the excess may itself be vacuum noise admitted by losses in the readout chain, which is quantum in origin, and assigning it to the classical side can only understate the quantum share. Writing $\nu = \nu_{\mathrm{SQL}} + \nu_{\mathrm{exc}}$, with $\nu_{\mathrm{exc}}$ that excess, gives
\begin{equation}
\beta = \frac{1+\nu_{\mathrm{SQL}}}{2\bar{n}_{\mathrm{th}}+1+\nu_{\mathrm{SQL}}+\nu_{\mathrm{exc}}},
\label{eq:beta_readout}
\end{equation}
and only $\nu_{\mathrm{exc}}$ degrades the ceiling. The terms of Eq.~(\ref{eq:beta_readout}) correspond to the kinds of noise distinguished in quantum measurement theory~\cite{Clerk2010}. The bath fluctuations are the fluctuation--dissipation force noise of the damping channel, thermal in $2\bar{n}_{\mathrm{th}}$ and zero-point in the unit. They are intrinsic to the damped mode and are present whether or not it is measured. The meter adds imprecision, noise in its output, and back-action, a force it exerts on the mode; quantum mechanics bounds the product of the two from below, and a linear position detector at its optimal coupling, where the two contribute equally, adds on resonance exactly the zero-point noise of the oscillator~\cite{Clerk2010}. This is again $\nu_{\mathrm{SQL}} = 1$, obtained for a position detector rather than for an amplifier, and it identifies the unit of Eq.~(\ref{eq:nu_sql}) as noise of the meter. The remainder, $\nu_{\mathrm{exc}}$, is the technical excess of the chain. The numerator of $\beta$ collects the quantum-origin terms of the first two kinds, the bath's zero-point unit and $\nu_{\mathrm{SQL}}$; which of them can be removed is a separate question, taken up term by term in Sec.~\ref{sec:ceiling}. The condition $\beta > 1/2$ becomes
\begin{equation}
2\bar{n}_{\mathrm{th}} + \nu_{\mathrm{exc}} < 1 + \nu_{\mathrm{SQL}},
\label{eq:joint}
\end{equation}
a joint requirement on the temperature of the mode and on the quality of the measurement that reads it. For a phase-sensitive readout it has the form of the mode's own condition with the excess added to the occupation. For a phase-insensitive readout at its limit it reads $\bar{n}_{\mathrm{th}} < 1$ rather than $\bar{n}_{\mathrm{th}} < 1/2$, because the added half-quantum is itself quantum in origin. The thermal frontier of Sec.~\ref{sec:frontier} remains a property of the mode, and Eq.~(\ref{eq:joint}) is the corresponding statement for the instrument. The larger quantum share it allows is not a larger removable share, which for the same readout is capped at one half (Sec.~\ref{sec:ceiling}): every term of the denominator has an origin and a removability, and $\beta$ reads only the first. For intrinsic damping the bath's zero-point unit will prove unremovable, and retaining only $\nu_{\mathrm{SQL}}$ in the numerator will define the addressable fraction $\beta_a$ of Sec.~\ref{sec:ceiling}. Sections~\ref{sec:frontier}--\ref{sec:proposal} ask where quantum noise dominates and what crossing the frontier costs, questions of origin alone, for which $\beta$ is the right variable. Equation~(\ref{eq:joint}) also serves as a design specification that constrains only the excess: with a phase-insensitive readout whose excess reaches two zero-point units, the instrument is not quantum dominated however deeply its mode is cooled.

\section{The thermal frontier}
\label{sec:frontier}

Setting $\bar{n}_{\mathrm{th}} = 1/2$ in Eq.~(\ref{eq:beta_mech}) defines a boundary in the frequency--temperature plane,
\begin{equation}
T < T_q(f) \equiv \frac{\hbar\omega_0}{k_B\ln 3} = 43.7\;\mathrm{mK}\times\left(\frac{f}{1\;\mathrm{GHz}}\right),
\label{eq:Tq}
\end{equation}
which we call the \emph{thermal frontier}. The logarithm comes from Eq.~(\ref{eq:beta_mech}): the frontier is $\tanh u = 1/2$, so $u = \operatorname{artanh}(1/2) = \tfrac12\ln 3$. The other natural boundaries are markers of the same variable, differing from this one and from one another: mean occupation below unity is $u = \tfrac12\ln 2$, where $\beta = 1/3$; $\hbar\omega \ge 2k_BT$ is $u = 1$ (the zero-point energy $\hbar\omega/2$ overtaking $k_BT$, the most natural threshold for a dimensionless argument), where $\beta = 0.762$. The equal-share condition lies between, at $u = \tfrac12\ln 3$. Quantitatively, the frontier temperature at fixed frequency is 37\% below the $\bar{n}_{\mathrm{th}} = 1$ criterion, and the frontier frequency at fixed temperature 45\% below the $\hbar\omega = 2k_BT$ criterion. On the cold side of the frontier, $T < T_q(f)$ or, at fixed temperature, above the frontier frequency, $\beta > 1/2$ and $\mathcal{E}_{\mathrm{max}} > 1.41$; on the warm side the quantum share falls off, and the return on quantum technique with it.

\begin{figure*}
\includegraphics[width=0.75\textwidth]{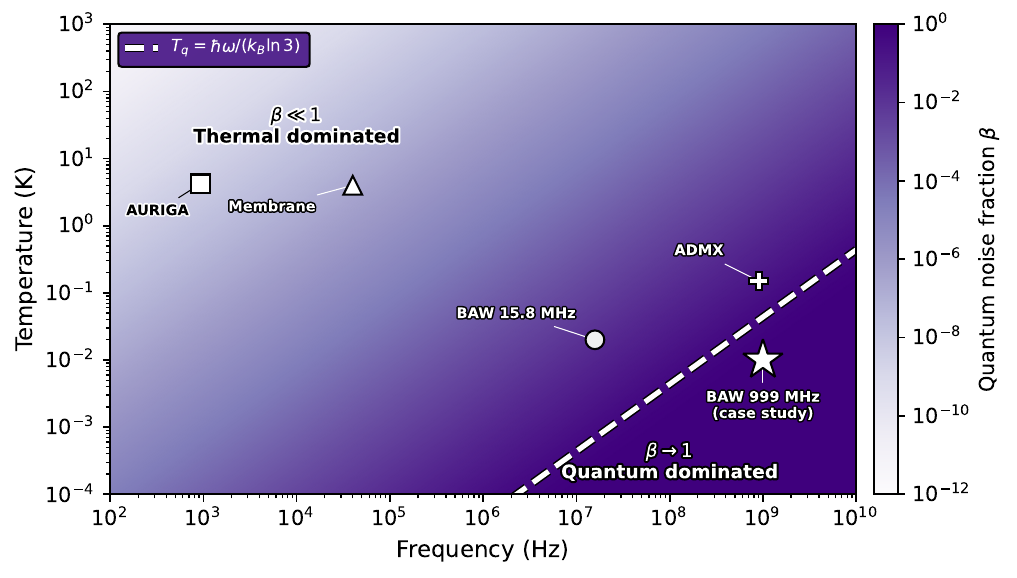}
\caption{\label{fig:frontier}Quantum noise fraction $\beta$ in the frequency--temperature plane. The dashed line marks the thermal frontier $T_q = \hbar\omega/(k_B\ln 3)$: below it $\beta > 1/2$ and the enhancement ceiling exceeds $\sqrt{2}$; above it thermal occupation dominates the budget. Markers show representative detectors at their operating points.}
\end{figure*}

At representative frequencies $T_q$ equals 44~nK at 1~kHz, 44~$\mu$K at 1~MHz and 43.7~mK at 1~GHz. Only the last is experimentally accessible for a macroscopic system, through dilution refrigeration to $\sim$10~mK, which is routine in superconducting circuit laboratories. Equivalently, at $T = 10$~mK the frontier lies at $f = 229$~MHz. Figure~\ref{fig:frontier} maps $\beta$ across the plane with representative detectors marked.

The scope of Eq.~(\ref{eq:Tq}) is broader than the mechanical case that motivates it, since its only inputs are a signal registered in a bosonic mode of frequency $\omega_0$ and a bath at temperature $T$. A Mechanism~D detector has the same structure: the wave is converted, in the static field, into a signal accumulating in a resonant cavity mode, and the competing noise is the blackbody occupation of that same mode at the signal frequency. Equations~(\ref{eq:beta_mech}) and~(\ref{eq:Tq}) apply unchanged, and the frontier lies at the same place for a microwave cavity as for an acoustic resonator. The two mechanisms sit on opposite sides of the companion's direct/indirect divide, one deforming a crystal through the tidal field and the other converting the wave into photons, and their coupling strengths depend on unrelated quantities (mode mass, coupling length and quality factor in one case; field strength, cavity volume and geometric overlap in the other), none of which enters $\beta$. The qualitative statement that, at about 10~GHz, blackbody occupation dominates the microwave noise budget unless the system is cooled to sub-kelvin temperatures has been noted in the literature~\cite{CaiVisinelliYan2025}; Eq.~(\ref{eq:Tq}) makes it exact and ties it to the enhancement ceiling. Mechanism~C is the exception, for a reason that follows from the same argument: an optical mode at 1064~nm has $\bar{n}_{\mathrm{th}} \sim 10^{-20}$ even at room temperature, so an interferometer never meets a thermal frontier, and its $\beta$, set by the ratio of shot noise to the classical floor, sits at 0.85--0.95 while resonant masses and microwave cavities on the warm side of the frontier sit well below one half.

For the existing landscape the consequence is that every detector whose signal is registered in a mechanical mode well below $\sim$230~MHz, operating or proposed, has $\beta \ll 1$ (Table~\ref{tab:landscape}), whether dilution-cooled or not. This includes cryogenic bars, tested down to about 140~mK~\cite{AURIGA} and operated at liquid-helium temperature~\cite{Branca2017}, BAW phonon-trapping resonators at MHz frequencies~\cite{Goryachev2014,Goryachev2021}, optomechanical membranes at 0.5--40~kHz in a 4~K cryostat~\cite{Membrane2025}, and levitated sensors at 50--300~kHz~\cite{Arvanitaki2013}. Mechanical squeezing, back-action evasion and entangled readout remain valuable there for state preparation and tests of quantum mechanics, but their return in strain sensitivity is bounded by Eq.~(\ref{eq:enhancement}) and correspondingly small.

\begin{table}
\linespread{1}\selectfont
\caption{\label{tab:landscape}Quantum noise fraction $\beta = 1/(2\bar{n}_{\mathrm{th}}+1)$ and strain sensitivity for representative high-frequency detectors. For Mechanisms~A and~D, $\beta$ follows from the Bose--Einstein occupation at the operating frequency and temperature, excluding the readout [cf.\ Eq.~(\ref{eq:beta_readout})]; for Mechanism~C it is the shot-noise fraction of the budget~\cite{Gaudio2026CQG}. The AURIGA entry is the achieved sensitivity of Ref.~\cite{Branca2017}, placed at the 930~Hz to which its transducer circuit is tuned, with the bar at liquid-helium temperature, taken here as 4.2~K; the membrane entry is the peak sensitivity projected in Ref.~\cite{Membrane2025}, at 40~kHz and its assumed 4~K. The membrane is driven by the radiation-pressure force of its optical trap, whose position the wave modulates through the length of a 100~m cavity~\cite{Membrane2025}; its $\beta$ is that of its mechanical mode. The LIGO entry is read from Fig.~2 of Ref.~\cite{Jia2024}, which states no numerical sensitivity (reading uncertainty $\sim$20\%), and is placed at 300~Hz, where the $\beta = 0.85$ of Ref.~\cite{Gaudio2026CQG}, quoted there for $f \gtrsim 200$~Hz, applies; the ET-HF entry is the high-frequency interferometer of the ET-D configuration of Ref.~\cite{Hild2011} at the same frequency, read from the official ET-D sensitivity file~\cite{ETD0000A18}, in which the low-frequency interferometer is some 300 times less sensitive at 300~Hz, so that the ET-D total coincides there with the high-frequency interferometer, whose $\beta$ is the one quoted for ET-HF in Ref.~\cite{Gaudio2026CQG}. The ADMX entry is a dimensionless strain reach, reported as Ref.~\cite{Berlin2022} quotes it rather than converted to a spectral density; its $\beta$, evaluated at 900~MHz and at the 150~mK cavity temperature of the 3.3--4.2~$\mu$eV ADMX run, 798--1016~MHz~\cite{Bartram2021} (0.13--0.16 across that band), would become 0.250 with a phase-insensitive readout at the standard quantum limit (higher, because the added half-quantum is itself quantum in origin), 0.200 at $\nu = 3$, and 0.072 at the system noise temperature reported for that run, $\nu \simeq 20.8$ (Sec.~\ref{sec:ceiling}). Both BAW entries follow from Eq.~(\ref{eq:Sh}) for the device of Ref.~\cite{Goryachev2014} with trapping parameter $\eta_{\mathrm{tr}} = 1$ (Sec.~\ref{sec:knee}): the fifth overtone with the $Q = 10^{9}$ projected, not measured, for it at 20~mK, and the 317th with $Q = 10^{8}$, an extrapolation discussed in Sec.~\ref{sec:proposal}. The entry in bold is the case study of Sec.~\ref{sec:proposal}.}
\begin{ruledtabular}
\begin{tabular}{lccc}
Detector & $f$ & $\sqrt{S_h}$ [Hz$^{-1/2}$] & $\beta$ \\
\hline
\multicolumn{4}{l}{\emph{Mechanism A (tidal coupling to a mechanical mode) and other mechanical resonators}} \\
AURIGA, 4.2~K & 930~Hz & $2\times 10^{-21}$ & $5.3\times 10^{-9}$ \\
Membrane, 4~K & 40~kHz & $2\times 10^{-23}$ & $2.4\times 10^{-7}$ \\
BAW quartz, 20~mK & 15.8~MHz & $2.2\times 10^{-22}$ & $1.9\times 10^{-2}$ \\
\textbf{BAW quartz, 10~mK} & \textbf{999~MHz} & $\bm{4.8\times 10^{-20}}$ & \textbf{0.98} \\[3pt]
\multicolumn{4}{l}{\emph{Mechanism C (light propagation)}} \\
LIGO O4, squeezed~\cite{Jia2024} & 300~Hz & $3.7\times 10^{-24}$ & 0.85 \\
ET-HF (high-frequency interferometer), projected~\cite{Hild2011} & 300~Hz & $3.2\times 10^{-25}$ & 0.95 \\[3pt]
\multicolumn{4}{l}{\emph{Mechanism D (conversion in a static $B$ field)}} \\
ADMX cavity~\cite{Berlin2022} & 900~MHz & $h \sim 10^{-22}$--$10^{-21}$ & 0.14 \\
\end{tabular}
\end{ruledtabular}
\end{table}

\section{Tidal coupling of an acoustic mode}
\label{sec:BAW}

The frontier says where quantum noise dominates a budget; what crossing it buys can only be judged on a concrete device. We take the bulk acoustic wave resonator, for two reasons. It is the platform on which phonon trapping, the confinement of the acoustic mode about the axis of a contoured plate, has been demonstrated up to 712~MHz in a macroscopic crystal (Sec.~\ref{sec:proposal}), which makes it the one resonant mass that can operate on the cold side of the frontier at dilution temperatures. Its spectrum is an overtone comb spanning up to three decades on one crystal, and this is what makes the frontier observable within a single device (Sec.~\ref{sec:knee}). The quantity required is its strain sensitivity. The derivation is elementary, but at one step an incorrect choice of mode basis produces an error by a factor of several hundred in the coupling for this device, and more for a thicker plate.

A crystalline plate of thickness $L$, cross-section $A$ and mass $M = \rho A L$ supports thickness modes at $f_n = n v_s/(2L)$. The surfaces of a free plate are traction-free, so the boundary condition on the displacement is $\partial u_n/\partial z = 0$ at both faces and the mode functions are $u_n(z) = \cos(n\pi z/L)$ on $z \in [0,L]$. The distinction from the clamped basis $\sin(n\pi z/L)$, which vanishes at the faces, is not a matter of convention. A gravitational wave enters through the tidal acceleration $a(z) = \tfrac{1}{2}\omega^2 h\,z$ in the local proper frame, and the position of the coordinate origin is a gauge choice: shifting it adds a uniform acceleration, which is unobservable. The free-plate modes satisfy $\int_0^L u_n\,dz = 0$ for every $n \ge 1$, so they are orthogonal to rigid translation and the overlap with a uniform acceleration vanishes identically; the coupling is therefore independent of where the origin is placed, as the equivalence principle demands. The clamped basis has $\int_0^L u_n\,dz \ne 0$ for odd $n$ and yields a coupling that depends on the origin, which is itself the signal that the basis is inappropriate to a free crystal.

With the correct basis the overlap integral is
\begin{equation}
I_n = \int_0^L z\cos\!\left(\frac{n\pi z}{L}\right)dz = \frac{[(-1)^n - 1]\,L^2}{n^2\pi^2},
\label{eq:overlap}
\end{equation}
so that $I_n = 0$ for even $n$ and $|I_n| = 2L^2/(n^2\pi^2)$ for odd $n$. Only odd overtones couple to the tidal field. This selection rule has the same origin, the antisymmetry of the displacement about the center of the plate, as the experimental fact that only odd overtones of these devices can be excited piezoelectrically~\cite{Galliou2013,Goryachev2014}, so the modes accessible to the readout are the modes that respond to the wave.

Defining the effective coupling length through $F_n = \tfrac{1}{2}M_{\mathrm{eff}}\omega^2 h\,\ell_n$ with $M_{\mathrm{eff}} = M/2$ gives
\begin{equation}
\ell_n = \frac{2|I_n|}{L} = \frac{4L}{n^2\pi^2} = \frac{\lambda_n^2}{\pi^2 L},
\qquad \lambda_n = \frac{2L}{n}.
\label{eq:ell}
\end{equation}
The second form is the more instructive: the coupling length is smaller than the acoustic wavelength by a further factor $\lambda_n/L$, so that at high overtone it is suppressed quadratically rather than linearly in $n$. For the 1~mm plate of Sec.~\ref{sec:proposal}, operating at 1~GHz on its 317th overtone, this is a suppression by $n\pi/2 = 498$ relative to $\lambda_n/\pi$; a thicker plate, needing a higher overtone to reach the same frequency, fares worse still.

Equation~(\ref{eq:ell}) agrees with the corrected coupling coefficient of Goryachev and Tobar~\cite{Goryachev2014}: evaluating their overlap integral for a flat, untrapped plate returns $8h_0/(n^2\pi^2)$ with $L = 2h_0$, which coincides with Eq.~(\ref{eq:ell}), while tight trapping gives twice that; the $n^{-2}$ scaling is common to both limits.

In the companion's classification the plate is a Mechanism~A detector. The operator is the one defining Mechanism~A, the tidal Hamiltonian $-\tfrac{1}{4}\ddot{h}_{ij}\int\rho\,x_ix_j\,d^3x$, projected onto an elastic mode rather than an electronic coordinate, and the orthogonality argument above shows that the coupled coordinate is internal rather than the center of mass: the rigid-translation component produces no excitation, and what remains is a deformation.

The difference from the companion's atomic case lies in the detection channel rather than in the coupling, and it matters for that paper's no-go result. There the quantity of interest is the energy shift of discrete internal states, which vanishes at first order for the $J = 0$ clock states and is unobservably small at second order; here nothing is asked of an internal level. The wave exerts a classical generalized force on a harmonic mode, the mode responds classically, and quantum mechanics enters only in the noise against which the response is measured. The coupling strengths differ greatly, the product of mode mass and coupling length fixing the tidal force, $m_\lambda\xi$ for the mode of Sec.~\ref{sec:proposal} with the mode function normalized to unit maximum as in Ref.~\cite{Goryachev2014}, exceeding the atomic $m_ea_0$ by $2.5\times10^{26}$, almost entirely through the mass. What exempts a resonant mass from the no-go result, however, is the change of observable and not the change of scale. It is a force detector with a quantum-limited noise floor, which is the regime in which $\beta$ and $\mathcal{E}_{\mathrm{max}}$ are the relevant figures of merit.

The strain noise follows from the quantum fluctuation--dissipation theorem. The single-sided symmetrized force noise of a damped oscillator is $S_F = 2M_{\mathrm{eff}}\gamma\hbar\omega_0(2\bar{n}_{\mathrm{th}}+1)$ with $\gamma = \omega_0/Q$; combining this with the on-resonance susceptibility $|\chi(\omega_0)|^2 = Q^2/(M_{\mathrm{eff}}^2\omega_0^4)$ and the strain-to-displacement relation $h = 2x/(\ell_n Q)$ gives
\begin{equation}
S_h = \frac{8\,\hbar}{M_{\mathrm{eff}}\,\omega_0^2\,\ell_n^2\,Q}\,(2\bar{n}_{\mathrm{th}} + 1),
\label{eq:Sh}
\end{equation}
which separates as $S_h^{\mathrm{zpf}} = 8\hbar/(M_{\mathrm{eff}}\omega_0^2\ell_n^2 Q)$ and $S_h^{\mathrm{th}} = 2\bar{n}_{\mathrm{th}}S_h^{\mathrm{zpf}}$. The ratio $S_h^{\mathrm{th}}/S_h^{\mathrm{zpf}} = 2\bar{n}_{\mathrm{th}}$ reproduces Eq.~(\ref{eq:beta_mech}), as it must. Equation~(\ref{eq:Sh}) is algebraically identical to the resonant-mass expression of Ref.~\cite{Goryachev2014} under the correspondence $\ell_n \leftrightarrow \xi_\lambda$, $M_{\mathrm{eff}} \leftrightarrow m_\lambda$, where $\xi_\lambda$ and $m_\lambda$ are the coupling length and mode mass of a trapped mode; in the untrapped limit $\xi_\lambda$ reduces to $\ell_n$, and $m_\lambda$ to $M/2$ up to a factor $4/\pi$ that comes from the square integration domain of that reference; we have verified the identity numerically. In the classical limit $k_BT \gg \hbar\omega_0$ it reduces to $S_h \to 16k_BT/(M_{\mathrm{eff}}\omega_0^3\ell_n^2 Q)$, which for $n = 1$, with $\ell_1 = 4L/\pi^2$ and $M_{\mathrm{eff}}$ the effective mass of the bar, is the square of the thermal-noise strain sensitivity of a resonant bar of Ref.~\cite{TobarBlair1995} in its corrected form.

\section{Overtone scaling across the frontier}
\label{sec:knee}

A bulk acoustic wave device is not a single-mode instrument: its spectrum is a comb of odd overtones spanning up to three decades, sharing one crystal, one cryostat and one readout chain, and the thermal frontier crosses the comb at a definite place that can be measured within one device, which no comparison between instruments allows.

The frequency dependence of the sensitivity along the comb follows from Eq.~(\ref{eq:Sh}) once the trapped mode structure is included. For a contoured plate the mode mass falls with overtone number as $m_\lambda \propto n^{-1}$~\cite{Goryachev2014}, while the trapped coupling length falls as $\xi \propto n^{-2}$ (the same power as the flat-plate $\ell_n$ of Eq.~(\ref{eq:ell}), since the error functions of the trapping integrals saturate at high overtone), and $f_n \propto n$. The trapped pair $(m_\lambda,\xi)$ is used here and in the case study of Sec.~\ref{sec:proposal}, because Ref.~\cite{Goryachev2014} evaluates that plate with the trapped-mode expressions of a contoured cavity; its two members go together, since it is under their joint correspondence with $M_{\mathrm{eff}}$ and $\ell_n$ that Eq.~(\ref{eq:Sh}) reproduces the expression of that reference. Writing $\sqrt{S_h} \propto (f_n\xi)^{-1}m_\lambda^{-1/2}(\bar{n}_{\mathrm{th}}+\tfrac12)^{1/2}$ and collecting powers,
\begin{equation}
\sqrt{S_h} \;\propto\; n^{3/2}\left(\bar{n}_{\mathrm{th}} + \tfrac{1}{2}\right)^{1/2}.
\label{eq:scaling}
\end{equation}
The bracket carries the whole of the temperature dependence. Well below the frontier frequency the occupation is classical, $\bar{n}_{\mathrm{th}} \simeq k_BT/\hbar\omega_n \propto n^{-1}$, the bracket contributes $n^{-1/2}$, and the sensitivity degrades linearly with overtone number. Well above it the occupation is negligible, the bracket saturates at $1/\sqrt2$, and the degradation steepens to the three-halves power:
\begin{equation}
\sqrt{S_h} \propto n \quad (T \gg T_q), \qquad
\sqrt{S_h} \propto n^{3/2} \quad (T \ll T_q).
\label{eq:knee}
\end{equation}
The crossover is thus not only a criterion classifying detectors but a change in the scaling law of a single one, and every point of it, the frontier included, moves in proportion to temperature, as Eq.~(\ref{eq:Tq}) prescribes.

\begin{figure*}
\includegraphics[width=0.9\textwidth]{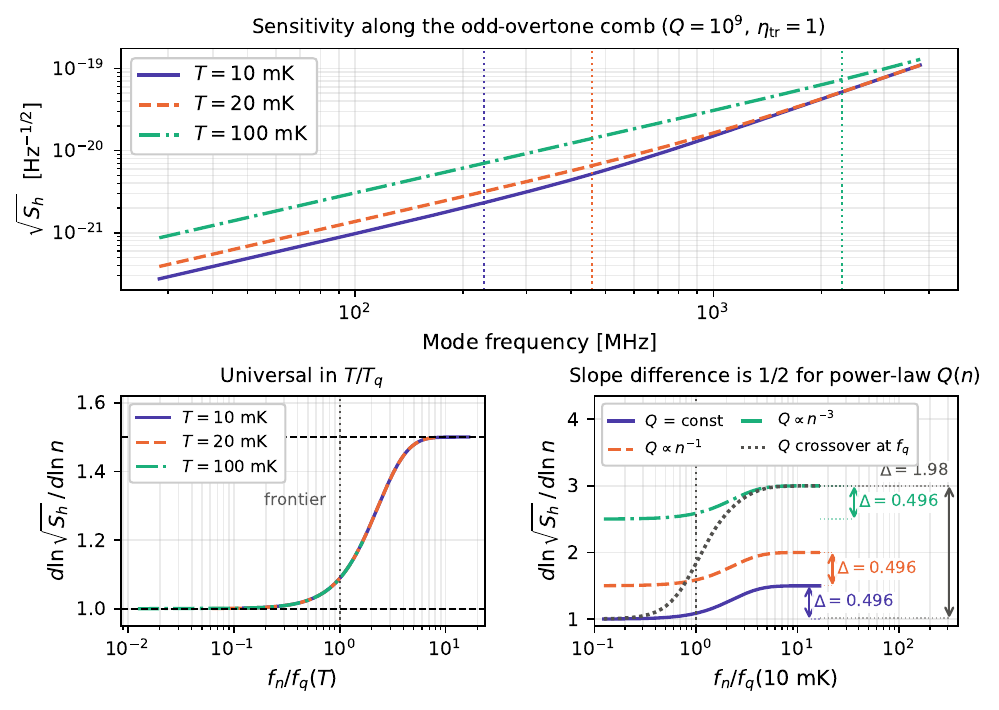}
\caption{\label{fig:knee}The overtone knee and its invariance. \textbf{Upper panel:} strain sensitivity of successive odd overtones of a single BAW plate at three temperatures, for the device parameters of Ref.~\cite{Goryachev2014} with $Q = 10^{9}$ held fixed along the comb, the value projected, not measured, for that device at 20~mK, and trapping parameter $\eta_{\mathrm{tr}} = 1$; the case study of Sec.~\ref{sec:proposal} adopts $Q = 10^{8}$, extrapolated to GHz frequencies, which raises the curve by $\sqrt{10}$ without altering its shape. Dotted lines mark the thermal frontier at each temperature. \textbf{Lower left:} the local logarithmic slope $d\ln\sqrt{S_h}/d\ln n$ against frequency in units of each curve's own frontier frequency; the three temperatures collapse onto one curve, which is the universality in $T/T_q$. \textbf{Lower right:} the same slope for three assumed power-law dependences of the quality factor, $Q \propto n^{-p}$ with $p = 0$, 1 and 3. The absolute slopes are displaced by $p/2$, but the change across the frontier is the same in every case. The dotted gray curve shows why the power-law assumption cannot be taken for granted: for a quality factor with a plateau-to-Rayleigh crossover placed at the frontier, $Q^{-1} = Q_0^{-1}[1+(n/n_q)^3]$, the same construction returns 1.98 rather than $1/2$. Forming $\sqrt{S_hQ_n}$ with the measured linewidths removes this contamination exactly, which is why Eq.~(\ref{eq:invariant}) is written with $Q$ inside the logarithm. All panels are drawn for $n \ge 9$, below which, at $\eta_{\mathrm{tr}} = 1$, the trapping integrals of Ref.~\cite{Goryachev2014} are not yet saturated and the slope departs from unity for reasons unrelated to thermal occupation.}
\end{figure*}

Figure~\ref{fig:knee} shows the sensitivity, its local logarithmic slope, and that slope under a frequency-dependent quality factor, for the device of Ref.~\cite{Goryachev2014} at three temperatures, from Eq.~(\ref{eq:Sh}) with the trapped mass and coupling of that reference. The slope stays within 4\% of unity to about half the frontier frequency, passes through 1.09 at the frontier, and reaches $3/2$ to within 0.1\% a decade above. The crossover is continuous, and the frontier is a marker on it rather than its center: the slope reaches its midpoint, $5/4$, at about twice the frontier frequency. At $T = 100$~mK the frontier lies at 2.3~GHz and the slope reaches only 1.14 by 3~GHz: the effect is confined to low temperatures.

The prediction does not depend on a model because it rests on an invariant, the difference between the two exponents rather than the exponents themselves. The slope depends on frequency and temperature only through $\hbar\omega/k_BT$, so it is a universal function of $T/T_q$: at a given multiple of the frontier frequency it is the same at every temperature. Writing $f_q(T) = k_BT\ln 3/(2\pi\hbar)$, which is Eq.~(\ref{eq:Tq}) read as a frequency at fixed temperature, the three temperature curves collapse onto one when plotted against $f/f_q(T)$, as the lower left panel of Fig.~\ref{fig:knee} shows. Nor does the trapping parameter of Ref.~\cite{Goryachev2014} appear, the dimensionless number fixing how tightly the contoured plate traps the phonon about its axis: once the error functions of the trapping integrals saturate, the trapped mode mass falls as $\eta_{\mathrm{tr}}^{-2}$ while the coupling length is untouched. That reference writes it $\eta_x = \eta_y$; we add the subscript to keep it apart from the noise-reduction factors $\eta_i$ of Eq.~(\ref{eq:enhancement}), and we avoid its $\alpha$, which there denotes the Gaussian trapping coefficient. Entering the sensitivity as a constant multiplicative factor, it cancels from a logarithmic derivative; we have verified this over $\eta_{\mathrm{tr}} = 0.7$--3 (for smaller values the trapping integrals saturate only above the ninth overtone), and the case study of Sec.~\ref{sec:proposal} takes $\eta_{\mathrm{tr}} = 1$. The material constants and the plate geometry are common to all points of one crystal and cancel for the same reason.

The quality factor needs more care. Equation~(\ref{eq:knee}) holds at fixed $Q$, but $Q$ is not constant along the comb: at the highest overtones the observed $Q\times f$ product is limited by Rayleigh-type scattering from dilute lattice impurities~\cite{Goryachev2013}, so the measured exponents are 1 and $3/2$ displaced by whatever $Q(n)$ contributes. What survives is the \emph{change} in slope across the frontier, which is stated cleanly with the quality factor inside the logarithm:
\begin{equation}
\Delta\left(\frac{d\ln\sqrt{S_h\,Q}}{d\ln n}\right)_{\text{across the frontier}} \;=\; \frac{1}{2}.
\label{eq:invariant}
\end{equation}
In this form the result holds for any $Q(n)$ whatever, since dividing the measured quality factor out removes its contribution to the slope at each point separately, leaving the bracket of Eq.~(\ref{eq:scaling}) as the only frequency-dependent factor.

For a power law, $Q \propto n^{-p}$, $\sqrt{S_h}$ alone would suffice: the factor $n^{p/2}$ displaces the slope by $p/2$ in \emph{both} regimes and cancels from the difference, and the lower right panel of Fig.~\ref{fig:knee}, sampling at one fifth and at ten times the frontier frequency, returns 0.496 for $p = 0$, 1 and 3 alike, the small shortfall against $1/2$ being the finite separation of the sampling points, not a residual dependence on $p$.

That cancellation is fragile, which is why Eq.~(\ref{eq:invariant}) carries $Q$. It requires $d\ln Q/d\ln n$ to take the \emph{same value at both sampling points}, which a power law guarantees and a real $Q(n)$ need not: the quality factor is expected to hold a plateau and then fall as Rayleigh scattering sets in, at a position not known a priori. In quartz at 3.8~K the quality factor peaks at 100--200~MHz and falls as $f^{-3}$ above, which Ref.~\cite{Goryachev2013} attributes to Rayleigh-type scattering, while the frontier for $T = 10$~mK lies at 229~MHz: the two are close. Modeling the crossover as $Q^{-1} = Q_0^{-1}[1 + (n/n_c)^3]$, with $n_c$ the overtone at which Rayleigh scattering overtakes the plateau (the cubic exponent is not a free choice: Rayleigh scattering gives $Q_{\mathrm{Rayleigh}} \propto f^{-3}$, the form observed at high overtones in quartz BAW resonators~\cite{Goryachev2013}), and varying its position, the slope difference of $\sqrt{S_h}$ alone is 0.66 for $n_c = 0.1\,n_q$, 1.65 for $n_c = 0.3\,n_q$ and 1.98 for $n_c = n_q$: with the crossover at the frontier the measurement returns 2 where the theory predicts $1/2$, an error of a factor four that would be read as a refutation.

Forming $\sqrt{S_h\,Q_n}$ with the measured $Q_n$ removes the problem exactly. Repeating the calculation above for $n_c/n_q = 0.3$, 1 and 3 returns 0.4958 in every case, the same value as for a constant $Q$. The $Q_n$ follow from the linewidths of the modes, which are measured in any case to identify them, so the correction uses data the experiment already has.

The apparatus exists: MAGE, a multimode BAW antenna built from two near-identical quartz resonators at 4~K, was taking data intermittently in early 2023, with a spectral sensitivity estimated at as low as $6.6\times 10^{-21}\,\mathrm{Hz}^{-1/2}$ in multiple narrow bands across MHz frequencies, and its authors plan an additional detector to identify strains localized on a single resonator~\cite{MAGE2023}. Extending it to overtones spanning the frontier, at a temperature that brings the frontier inside the accessible comb, is a change of operating point, not of technology.

Equation~(\ref{eq:invariant}) is therefore the form in which we propose the test: the difference between two logarithmic slopes on the same crystal, requiring no trapping parameter, no material constants, no plate geometry and no absolute strain calibration, only the measured $Q_n$ as numbers rather than as a model. It does assume that the transduction from the mode to the recorded signal is calibrated between the sampling points or varies as a power of $n$, that each mode is at the temperature of the bath, and that the readout noise does not itself depend on $n$ or $Q_n$, as it would for an imprecision fixed in displacement. The quantity to be reported is dimensionless. Between sampling points far below and far above the frontier it equals one half; between finite sampling points it takes the value of the same universal curve, 0.4958 between $f_q/5$ and $10f_q$, and 0.48--0.49, reached near 5~mK, between the thirteenth overtone of this plate, at 41~MHz, and the highest documented, at 712~MHz~\cite{Goryachev2013}, for any $\eta_{\mathrm{tr}}$ between 0.7 and 3 (from the ninth overtone the value would depend on $\eta_{\mathrm{tr}}$, since the trapping integrals are not yet saturated there). These values refer to the noise of the mode. A readout adding $\nu$ zero-point units enters the measured floors as $2\bar{n}_{\mathrm{th}}+1+\nu$: for $\nu$ independent of frequency the change far across the frontier is still one half, but the knee moves toward the joint condition of Eq.~(\ref{eq:joint}), and between the same two overtones, with a readout at the standard quantum limit, the value is 0.40--0.41 near 5~mK and at most 0.41--0.42, reached near 7~mK. The value of the mode is recovered by subtracting a calibrated $\nu$, or in part by separating the resonant peak from the flat background of the readout's imprecision.

The practical consequence runs against the intuition that motivates going to GHz in the first place. Along the comb the quantum noise fraction increases monotonically, at 10~mK from $\beta = 0.008$ at the 3~MHz fundamental to $\beta = 0.984$ at 1~GHz, but the raw sensitivity degrades faster. Applying the ceiling of Eq.~(\ref{eq:enhancement}) mode by mode, holding $Q = 10^{9}$ across the comb so that the comparison isolates the scaling (the case study of Sec.~\ref{sec:proposal} adopts $Q = 10^{8}$ at GHz, shifting the GHz figure below and the curves of Fig.~\ref{fig:knee} by $\sqrt{10}$), the 1~GHz overtone reaches $\sqrt{S_h}/\mathcal{E}_{\mathrm{max}} = 1.9\times 10^{-21}\,\mathrm{Hz}^{-1/2}$ against $4.1\times 10^{-23}\,\mathrm{Hz}^{-1/2}$ for the unenhanced fundamental: a factor 47 worse, despite an enhancement ceiling of 7.8 against 1.00. The fundamental is the mode for which the trapping description of Ref.~\cite{Goryachev2014} is least secure (caption of Fig.~\ref{fig:knee}), so we repeat the comparison against modes with saturated trapping integrals: 12.6 against the fifth overtone at 15.8~MHz, 7.0 against the ninth at 28.4~MHz. The conclusion does not depend on the fundamental. Even with every quantum contribution removed, which Sec.~\ref{sec:ceiling} argues is not possible, the low-frequency end of the comb remains the more sensitive. A formal crossover exists, since $\mathcal{E}_{\mathrm{max}}$ grows exponentially in $\hbar\omega/k_BT$ while $\sqrt{S_h}$ grows only as $n^{3/2}$; for this device at 10~mK it falls near 3.4~GHz, well above the highest overtone for which quality factor and trapping are documented, at 712~MHz~\cite{Goryachev2013}, and we do not propose it as a design point.

Crossing the thermal frontier therefore costs more in classical sensitivity than it returns in quantum enhancement.

\section{A GHz acoustic mode as a case study}
\label{sec:proposal}

To place concrete numbers on the quantum-dominated regime we evaluate a mode of an existing device: the quartz plate characterized in Refs.~\cite{Galliou2013,Goryachev2014}, the 1~mm thick, 30~mm diameter electrode-separated cavity of Ref.~\cite{Galliou2013} (its cavity number 3, with higher-grade surface polishing). The effective velocity for its longitudinal thickness modes is fixed by the measured spectrum rather than assumed: Ref.~\cite{Galliou2013} reports the fifth overtone at 15.7~MHz and the sixty-fifth at 204~MHz, both giving a fundamental of 3.14~MHz and hence $v_s = 2Lf_1 \simeq 6280$~m/s. We use $v_s = 6303$~m/s, which reproduces those frequencies to 0.4\% and corresponds, with $\rho = 2643$~kg/m$^3$, to $\hat{c}_z \approx 105$~GPa. This differs by 7\% from the nominal bulk quasi-longitudinal velocity of 6757~m/s quoted there for a doubly rotated cut, which would put the fifth overtone at 16.9~MHz rather than the observed 15.7~MHz. The odd overtone nearest 1~GHz is $n = 317$, at $f = 999.0$~MHz, well above the sixty-fifth, the highest measured on this plate. With the trapped-mode expressions of Ref.~\cite{Goryachev2014} at $\eta_{\mathrm{tr}} = 1$ this mode has mode mass $m_\lambda = 1.47$~mg and coupling length $\xi = 8.07$~nm, the trapped counterparts of $M_{\mathrm{eff}}$ and $\ell_n$, which must be used as a pair. We take $Q = 10^8$ and $T = 10$~mK. The quality factor is an extrapolation. Phonon trapping in this device class has been demonstrated up to the 227th overtone, at 712~MHz, and at 15~mK up to 518~MHz with a maximum loaded $Q\times f$ of $4\times 10^{17}$, while at 3.8~K the loaded quality factor peaks at 100--200~MHz and falls as $f^{-3}$ above~\cite{Goryachev2013}. Continuing that law from the maximum $Q\times f$ measured at 3.8~K, $8.6\times 10^{17}$ at 172.7~MHz, gives about $2.6\times 10^{7}$ at 999~MHz, and at 15~mK the quality factors above 110~MHz are lower than at 3.8~K~\cite{Goryachev2013}. Our value therefore lies above the measured trend, by a factor of about four or more; even the largest quality factor measured at 15~mK among the modes above 216.65~MHz, $6.96\times 10^{8}$, would continue to only $9.7\times 10^{7}$ if it belonged to the highest of them, at 518~MHz. Nothing in $\beta$, $\mathcal{E}_{\mathrm{max}}$ or the frontier depends on it, since none of them contains $Q$. The sensitivity does, as $Q^{-1/2}$, but the bandwidth $f/Q$ available for integration, set by the linewidth for a readout matched to the mode, grows in compensation, so that the integrated figure of Sec.~\ref{sec:gap} scales only as $Q^{-1/4}$: a value of $10^7$ instead of $10^8$ would widen the final shortfall from $1.0\times 10^{10}$ to $1.8\times 10^{10}$, and a value of $4\times 10^{8}$ would narrow it only to $7.2\times 10^{9}$. The conclusion is insensitive to this assumption.

The thermal occupation is $\bar{n}_{\mathrm{th}} = 0.0083$, placing the mode deep in its ground state, with $\beta = 0.984$ and $\mathcal{E}_{\mathrm{max}} = 7.8$. The thermal frontier at this frequency is $T_q = 43.6$~mK, comfortably above the operating temperature. Equation~(\ref{eq:Sh}) gives
\begin{align}
\sqrt{S_h^{\mathrm{th}}} &= 6.1\times 10^{-21}\,\mathrm{Hz}^{-1/2}, \label{eq:Sh_th_num}\\
\sqrt{S_h^{\mathrm{zpf}}} &= 4.7\times 10^{-20}\,\mathrm{Hz}^{-1/2}, \label{eq:Sh_zpf_num}\\
\sqrt{S_h^{\mathrm{total}}} &= 4.8\times 10^{-20}\,\mathrm{Hz}^{-1/2}. \label{eq:Sh_tot_num}
\end{align}
Zero-point noise exceeds thermal noise by nearly an order of magnitude in amplitude and a factor 60 in power, which is the content of $\beta = 0.984$. These figures refer to the mode alone, without readout noise. A readout adding $\nu$ zero-point units raises them, and every sensitivity derived from them below, by $\sqrt{1+\nu/(2\bar{n}_{\mathrm{th}}+1)}$ on resonance, a factor 1.41 at the standard quantum limit and 2.06 and 4.63 for the readout chains of Sec.~\ref{sec:ceiling} ($\nu = 3.3$ and $\nu \simeq 20.8$). For the figures below that grant the ceiling $\mathcal{E}_{\mathrm{max}}$, which remove the quantum-origin noise and keep the classical, the relevant factor is instead $\sqrt{1+\nu_{\mathrm{exc}}/2\bar{n}_{\mathrm{th}}}$, with the excess booked classical as in Sec.~\ref{sec:ceiling}: 1 at the standard quantum limit, where the ceiling grows to 11.0 in compensation, and at most 11.8 and 34.5 for these chains, less to the extent that their excess is vacuum noise admitted by losses. Across the linewidth the factor is larger for a meter optimized at the mode frequency, since away from resonance the least noise added by a position detector with uncorrelated imprecision and back-action, $\hbar|\chi|$ for a mode of susceptibility $\chi$, exceeds the zero-point noise $\hbar\,\mathrm{Im}\,\chi$. These figures, like all quoted here, are for optimal orientation: the plate responds to $h_{ij}\hat{n}_i\hat{n}_j$ along its thickness axis $\hat{n}$, with squared antenna pattern averaging to $\langle F_+^2 + F_\times^2\rangle = \langle \sin^4\theta\rangle = 8/15$ over sky and polarization. An isotropic unpolarized background of one-sided spectral density $S_h$, which counts both polarizations, therefore reaches the plate's strain channel with density $(4/15)\,S_h$, half of $8/15$ because each polarization carries $S_h/2$; the same normalization gives $S_h/5$ for an interferometer with orthogonal arms~\cite{AllenRomano1999}. Against such a background the quoted sensitivities are worse by $\sqrt{15/4} = 1.94$. We have not folded this factor in: it is common to every configuration of the case study, and leaving it out favors the detector, since including it would raise the final shortfall of Sec.~\ref{sec:gap} from $1.0\times 10^{10}$ to $2.0\times 10^{10}$. An idealized flat plate of 5~mm thickness and diameter in the same quartz, treated in the untrapped limit with $M_{\mathrm{eff}} = M/2$ and $\ell_n$ from Eq.~(\ref{eq:ell}), gives $5.1\times 10^{-20}\,\mathrm{Hz}^{-1/2}$ at the same frequency, temperature and quality factor, within 7\%. The agreement should not be over-read: through the mode mass the trapped figure scales as $\eta_{\mathrm{tr}}$ (Sec.~\ref{sec:knee}), a parameter that cannot be measured directly, although it can be estimated from the spurious resonances of an overtone, and below unity the quality factor is degraded by clamping losses~\cite{Goryachev2014}; a different value would move every sensitivity of the case study, and the final shortfall of Sec.~\ref{sec:gap}, by the same factor. The trapping is also what makes the assumed quality factor plausible at all.

\emph{Readout.} The framework of Sec.~\ref{sec:beta} assumes a linear readout. A BAW plate allows one directly, since a piezoelectric crystal couples to an electronic circuit; Ref.~\cite{Goryachev2014} takes a near-quantum-limited SQUID amplifier as the first stage, and a Josephson parametric amplifier, as used by axion haloscopes in the same band~\cite{Bartram2021}, is an alternative. If a fraction $\epsilon$ of the signal power reaches an amplifier that adds $A$ quanta referred to its input, a cold loss admits $(1-\epsilon)/2$ quanta of vacuum (a loss at occupation $\bar{n}_L$ admits $(1-\epsilon)(\bar{n}_L+\tfrac12)$), and referring both contributions back to the mode gives, in the units of Eq.~(\ref{eq:total_readout}),
\begin{equation}
\nu = \frac{(1-\epsilon) + 2A}{\epsilon} ,
\label{eq:nu_eps}
\end{equation}
valid for $0 < \epsilon \le 1$, so that a quantum-limited amplifier, $A = 1/2$, yields $\nu = (2-\epsilon)/\epsilon$, and the standard quantum limit $\nu = 1$ requires $\epsilon \to 1$. The values of $\nu$ realized by existing chains are taken up in Sec.~\ref{sec:ceiling}. A different route couples the mode to a superconducting qubit through the same piezoelectric interaction~\cite{Chu2017}: in the strong dispersive regime the qubit resolves the phonon number distribution and parity of a BAW mode~\cite{vonLupke2022}, and phonon-number splitting has been resolved on a GHz nanomechanical mode as well~\cite{Arrangoiz2019}. This is a measurement of phonon number rather than of a quadrature. It lies outside the linear premise of Eqs.~(\ref{eq:total_readout})--(\ref{eq:joint}), and the bounds of Sec.~\ref{sec:ceiling}, derived for a linear chain, do not apply to it. Addressing one overtone at a time is not an obstacle: the odd overtones that couple are spaced by $2f_1 = 6.30$~MHz, against a linewidth $f/Q = 10$~Hz, and individual addressing of many acoustic modes of a BAW resonator has been demonstrated~\cite{Chu2017}.

\emph{Array scaling and integration.} An array of $N$ independent devices improves the strain sensitivity by $\sqrt{N}$. Taking $N = 10^4$,
\begin{equation}
\sqrt{S_h^{\mathrm{array}}} = 4.8\times 10^{-22}\,\mathrm{Hz}^{-1/2}.
\end{equation}
This figure, and the one-year figure derived from it below, are best cases. The $\sqrt{N}$ gain assumes that all devices see the same signal and have independent noise, and each of these assumptions sets a demanding condition. The signal must fall within every device's bandwidth, so the modes must coincide to within $f/Q = 10$~Hz and remain so over the whole integration time. This is a fractional matching of $1/Q = 10^{-8}$, which for a 1~mm plate corresponds to a thickness uniform to 10~pm, and we know of no tuning mechanism at 10~mK with the range and resolution needed to impose it. The array must also be compact compared with the gravitational wavelength, which is 30~cm at 999~MHz. The overlap between two co-aligned plates, equal to one when they coincide, falls to 0.92 and 0.78 at a separation of 5~cm along and across their axis, and to 0.72 and 0.27 at 10~cm. Ten thousand devices of 5~cm$^3$ fill a cube of 37~cm side. Combining the pairs with inverse-variance weights, which for identical devices scale as the square of each pair's overlap, in the large-noise approximation of Ref.~\cite{AllenRomano1999}, recovers a fraction $\langle\gamma^2\rangle^{1/4} = 0.50$ of the co-located gain, and the one-year figure below becomes $7.2\times 10^{-24}\,\mathrm{Hz}^{-1/2}$. Finally, the noise must be independent between devices. Disturbances common to a cryostat would violate this condition. Energy depositions in a BAW cavity have been observed as rare transients~\cite{Goryachev2021}, and ionizing impacts degrade superconducting qubits across a whole device~\cite{McEwen2022}. Non-stationary events of this kind can be vetoed, but any stationary common-mode component adds directly to the cross-correlated signal and mimics a stochastic background~\cite{AllenRomano1999}.

If the frequencies can be matched only in pairs, with each pair at its own frequency, the array is no longer coherent. Summing the pairwise signal-to-noise ratios in quadrature then gives an amplitude gain of $(N/2)^{1/4} = 8.4$ over a single pair, compared with $[N(N-1)/2]^{1/4} = 84$ for the coherent array, and the one-year figure becomes $3.6\times 10^{-23}\,\mathrm{Hz}^{-1/2}$. Distinct frequencies do allow frequency-domain multiplexing of the readout. A coherent array does not, and needs a readout line for each device. With the passive heat load measured for coaxial lines in a dilution refrigerator, 1.4~$\mu$W for 66 lines against a cooling power of about 5~$\mu$W at 13~mK at the still power used there, and 19~$\mu$W at 20~mK at increased circulation~\cite{Krinner2019}, ten thousand lines would exceed the capacity of the mixing chamber by a factor between about ten and forty. The mass of the array is not a limitation: the crystals amount to 7~L, or about 19~kg of quartz (Ref.~\cite{Goryachev2014} quotes below 5~cm$^3$ per device with its vacuum enclosure), a small fraction of the 742~kg of crystals in the CUORE cryostat, whose base temperature is about 10~mK~\cite{CUORE2019}. The array is therefore an optimistic bound rather than a design. Each of these conditions weakens it, but none changes the conclusion of Sec.~\ref{sec:gap}, which is reached with the bound granted in full.

For a stochastic background the observation time enters too, and the exponent needs care. Cross-correlating two detectors accumulates signal-to-noise in the \emph{power} spectral density as $\sqrt{2\Delta f\,T_{\mathrm{obs}}}$ for one-sided spectra~\cite{AllenRomano1999}, and summing the $N(N-1)/2$ pairs of the array in quadrature gives $\sqrt{N(N-1)\,\Delta f\,T_{\mathrm{obs}}}$, which for $N \gg 1$ is $N\sqrt{\Delta f\,T_{\mathrm{obs}}}$ ($\sqrt{\Delta f\,T_{\mathrm{obs}}} = 1.8\times 10^{4}$ for $\Delta f = f/Q = 10$~Hz over one year). The minimum detectable amplitude spectral density therefore improves on that of a single device by the factor $\sqrt{N}$ of the array scaling above and, in addition, by the fourth root of $\Delta f\,T_{\mathrm{obs}}$,
\begin{equation}
(\Delta f\,T_{\mathrm{obs}})^{1/4} = 133,
\label{eq:integration}
\end{equation}
giving $3.6\times 10^{-24}\,\mathrm{Hz}^{-1/2}$ for the array after one year, at a signal-to-noise ratio of one. This takes the integration bandwidth to be the linewidth. Referred to strain the noise of the bath is flat across the resonance, so a readout adding less noise away from resonance would widen the band. A detector with a large power gain adds at least the zero-point noise $\hbar|\mathrm{Im}\,\chi|$ at every frequency, and with a roughly constant correlation between imprecision and back-action cannot reach even that across the resonance~\cite{Clerk2010}; a readout optimized on resonance at the standard quantum limit makes the one-year figure 1.26 times worse. A back-action-evading measurement of one quadrature need not add noise~\cite{Caves1982,Clerk2010} and would widen the band toward the 6.30~MHz spacing of the overtones, improving the one-year figure by at most a factor of about 24, a single quadrature carrying half the independent samples of the band; no such readout has been demonstrated for a bulk acoustic mode. Applying in addition the ceiling $\mathcal{E}_{\mathrm{max}} = 7.8$, a bound rather than an attainable gain (Sec.~\ref{sec:ceiling}), would give $4.6\times 10^{-25}\,\mathrm{Hz}^{-1/2}$. We carry all three figures forward, since the first is what the instrument delivers, the second what a campaign delivers under the conditions just stated, and the third what no configuration of this kind can exceed at that bandwidth.

\section{Which noise a quantum technique can remove}
\label{sec:ceiling}

The case study places the mode deep in its ground state, and Eq.~(\ref{eq:beta_mech}) accordingly assigns almost all of its noise a quantum origin. It does not follow that almost all of it can be removed. Which term of Eq.~(\ref{eq:Sh}) is addressable is a separate question from which term is quantum, and the answer is less favorable than $\beta$ alone suggests.

The zero-point contribution $S_h^{\mathrm{zpf}}$ is not the zero-point fluctuation of the resonator's state but of the bath it is dissipatively coupled to: the $\hbar\omega/2$ carried by the fluctuation--dissipation force noise $S_F = 2M_{\mathrm{eff}}\gamma\hbar\omega_0(2\bar{n}_{\mathrm{th}}+1)$ (single-sided; cf.\ Ref.~\cite{Clerk2010}), delivered to both mechanical quadratures alike. Preparing the mode in a squeezed state does not reduce it. Squeezed mechanical states below the zero-point level have been produced by reservoir engineering~\cite{Wollman2015,Lecocq2015}, but squeezing the state of the oscillator does not alter the force the bath exerts on it. Back-action evasion does not reduce it either, since it acts on the meter and not on the bath. Reducing $S_h^{\mathrm{zpf}}$ would require engineering the damping channel itself, injecting squeezed vacuum into the port through which the mode loses energy. This is possible when the dominant loss is an accessible coupling, but not when it is intrinsic, as are the bulk and surface losses that set $Q$ in quartz at millikelvin temperatures. This sharpens a remark of the companion paper, which distinguishes the classical squeezing of thermal mechanical motion at $\bar{n}_{\mathrm{th}} \gg 1$ from the squeezing of optical vacuum in an interferometer readout, and identifies resonant masses near their ground state as the regime in which mechanical squeezing would become relevant~\cite{Gaudio2026CQG}. Our conclusion is that even in that regime mechanical squeezing does not help, because what remains once $\bar{n}_{\mathrm{th}} \to 0$ is the bath's zero-point force, which the state of the mode does not control.

In quantitative form, the two labels of Sec.~\ref{sec:beta}, origin and removability, can be read off term by term. Of the terms in the denominator of Eq.~(\ref{eq:beta_readout}), the thermal occupation $2\bar{n}_{\mathrm{th}}$ and the excess readout noise $\nu_{\mathrm{exc}}$ are classical and lie outside $\beta$ by construction; the zero-point term of the bath is quantum in origin but, for intrinsic damping, is not removable by any operation on the mode or on the meter; and $\nu_{\mathrm{SQL}}$ is both quantum in origin and removable, by squeezed-state injection or back-action evasion in the readout chain. We therefore define the \emph{addressable fraction}
\begin{equation}
\beta_a = \frac{\nu_{\mathrm{SQL}}}{2\bar{n}_{\mathrm{th}}+1+\nu_{\mathrm{SQL}}+\nu_{\mathrm{exc}}},
\qquad
\mathcal{E}_a = \frac{1}{\sqrt{1-\beta_a}},
\label{eq:beta_addressable}
\end{equation}
which differs from $\beta$ by the omission of the bath's zero-point contribution from the numerator, and which bounds what quantum technique can deliver in such a device. Two of its properties are easily misread.

The first is that $\beta_a = 0$ when $\nu = 0$, and equally for a phase-sensitive readout at its limit, which adds nothing of quantum origin. The figures quoted so far, $\beta = 0.984$ and $\mathcal{E}_{\mathrm{max}} = 7.8$, describe the mode without a meter, and then the entire quantum share belongs to the bath and nothing of quantum origin can be removed. A readout changes both fractions, and the comparison must be made at fixed assumptions about it. For a readout at the standard quantum limit ($\nu_{\mathrm{SQL}} = 1$, $\nu_{\mathrm{exc}} = 0$) the same mode has $\beta = 0.992$ and $\beta_a = 0.496$, so that
\begin{equation}
\mathcal{E}_{\mathrm{max}} = 11.0
\qquad\text{against}\qquad
\mathcal{E}_a = 1.41 ,
\end{equation}
and an excess equal to the zero-point level lowers $\mathcal{E}_a$ to 1.22 if it is technical or thermal noise rather than loss. The improvement available is of order forty percent, not the factor of eleven the ceiling suggests, and no phase-insensitive chain can do much better, since with its excess booked classical $\beta_a \le 1/(2\bar{n}_{\mathrm{th}}+2) \le 1/2$ and therefore $\mathcal{E}_a \le \sqrt{2}$ for any chain. For a stationary background of random phase the gain is smaller still: removing $\nu_{\mathrm{SQL}}$ requires a phase-sensitive measurement of one quadrature, which halves the number of independent samples, and the minimum detectable amplitude then improves by $[(2\bar{n}_{\mathrm{th}}+2)/(\sqrt{2}\,(2\bar{n}_{\mathrm{th}}+1))]^{1/2} = 1.18$ rather than 1.41. (That this $\beta_a$ and the finite-window slope difference of Fig.~\ref{fig:knee} both print as 0.496 is a coincidence: one is $1/(2\bar{n}_{\mathrm{th}}+2)$ at 999~MHz and 10~mK, the other the asymptotic $1/2$ eroded by the finite sampling separation.)

The second is that $\beta_a$ does not grow as the meter degrades through losses, although losses add noise that is quantum in origin. A loss between the mode and the amplifier admits vacuum noise, the $(1-\epsilon)/\epsilon$ zero-point units of Eq.~(\ref{eq:nu_eps}), and that noise belongs to the quantum share of $\beta$, which the conservative bookkeeping of Eq.~(\ref{eq:beta_readout}) therefore understates. In general, however, this noise cannot be removed. A squeezed quadrature of variance $V$ injected upstream of the loss reaches the amplifier with variance $\epsilon V + (1-\epsilon)/2$, in units in which the vacuum is $1/2$, so the vacuum admitted by the loss survives any squeezing injected before it. It could be removed only through the port by which it enters, and the insertion losses of transducers, cables and circulators provide no such port. This is the criterion applied above to the bath: noise of quantum origin is removable only through an accessible coupling. Booking loss vacuum as addressable would make $\beta_a$ tend to unity, and $\mathcal{E}_a$ diverge, as $\epsilon \to 0$, for a meter that measures nothing. The conservative bookkeeping keeps $\nu_{\mathrm{exc}}$, losses included, out of the numerator of Eq.~(\ref{eq:beta_addressable}). The amplifier's own half-quantum, however, reaches the mode through the same loss, as $1/\epsilon$ units of Eq.~(\ref{eq:nu_eps}), and a phase-sensitive amplifier would remove all of it: for an excess due to loss $\beta_a$ stays close to one half and $\mathcal{E}_a$ close to $\sqrt{2}$, while for an excess of technical or thermal origin $\beta_a$ falls as the chain degrades, as the value 1.22 quoted above shows.

The two fractions coincide only where the bath's zero-point force is itself addressable, which requires the dominant damping channel to be an engineered coupling rather than an intrinsic loss. An interferometer is the limiting case in the other direction: there the quantum-origin noise is shot noise and radiation pressure, both properties of the light and both accessible, so $\beta_a \to \beta$ and the ceiling of the companion paper is the operative one. For a resonant mass in quartz at millikelvin temperatures the improvements they bound differ by a factor of eight, and it is $\beta_a$ that a designer needs.

The ceiling itself depends on the readout. With $2\bar{n}_{\mathrm{th}} = 0.017$ and a phase-insensitive readout, Eq.~(\ref{eq:joint}) keeps the mode quantum dominated up to $\nu_{\mathrm{exc}} = 1.98$, but the ceiling falls well before that point, from 11.0 at the standard quantum limit to 2.7 at $\nu_{\mathrm{exc}} = 0.3$ and 1.7 at $\nu_{\mathrm{exc}} = 1$. The readout chains demonstrated in the GHz range add more excess noise than this. The ADMX run already used for Table~\ref{tab:landscape} reports a typical $T_{\mathrm{sys}}/\epsilon$ of 600~mK, $\epsilon$ being the transmission efficiency between the cavity and the amplifier, with the cavity at 150~mK and the amplifier stage at 120~mK~\cite{Bartram2021}; at 900~MHz this is $\nu \simeq 20.8$, and 18--23 across the 798--1016~MHz band, for a chain operated well above the temperature assumed here. HAYSTAC reached a total of 2.3 noise quanta at 5.6--5.8~GHz, with the cavity at 127~mK~\cite{Zhong2018}, which bounds its readout at $\nu \le 3.3$, a bound rather than a value because part of the excess is thermal noise from the cavity's tuning rod. Read through chains of these two qualities, taking the HAYSTAC bound as a value, the mode of the case study would have $\beta = 0.09$ and 0.46, and $\mathcal{E}_a = 1.02$ and 1.14, respectively, for an excess of technical or thermal origin. For the BAW the values of $\beta$ are best cases, since the piezoelectric transduction between the mode and the amplifier adds its own loss through the efficiency $\epsilon$ of Eq.~(\ref{eq:nu_eps}), which we have not found quantified for a linear readout of a GHz bulk acoustic mode. This stringent specification on the readout chain is, in our view, the binding practical constraint on operating a resonant mass in the quantum regime, and the mode occupation alone does not capture it. The estimates that follow therefore quote sensitivities without quantum enhancement and carry $\mathcal{E}_{\mathrm{max}}$ only as a bound.

In the deep quantum limit, as $\hbar\omega_0/k_BT$ grows, $\bar{n}_{\mathrm{th}}\to 0$ and Eq.~(\ref{eq:beta_mech}) gives $1-\beta \simeq 2\bar{n}_{\mathrm{th}}$, so that
\begin{equation}
\mathcal{E}_{\mathrm{max}} \simeq \frac{1}{\sqrt{2\bar{n}_{\mathrm{th}}}}
= \frac{1}{\sqrt{2}}\,e^{\hbar\omega_0/2k_BT} \longrightarrow \infty .
\label{eq:ceiling_divergence}
\end{equation}
At 1~GHz and 10~mK this is the modest 7.8 computed above; at 10~GHz and the same temperature $\bar{n}_{\mathrm{th}} = 1.4\times 10^{-21}$ and the ceiling exceeds $10^{10}$. The divergence is correct but carries no practical content: if all noise of quantum origin were removed from a mode that contains almost nothing else, nothing would be left.

A recent proposal reads a Mechanism~D cavity with an entangled register of transmon qubits, reporting, for an idealized register of 800 qubits, a strain sensitivity five orders of magnitude beyond standard cavity-power limits~\cite{Yi2026}. This does not contradict Eq.~(\ref{eq:enhancement}): at 5.3~GHz and an assumed 20~mK, Eq.~(\ref{eq:ceiling_divergence}) already permits a factor 408, and the gain is in any case made against the excess readout noise, the $\nu_{\mathrm{exc}}$ of Sec.~\ref{sec:beta}, which is booked classical and which $\beta$ never bounded. The framework says where such gains must come from: in a cavity far on the cold side of the frontier thermal noise is negligible, and the only large term that can still be reduced is the excess noise of the readout.

\section{Origin of the sensitivity gap}
\label{sec:gap}

The strongest model-independent constraint on stochastic backgrounds at GHz frequencies comes from big bang nucleosynthesis (BBN), through the contribution of gravitational radiation to the relativistic energy density at that epoch. The joint BBN and cosmic microwave background analysis of Ref.~\cite{Yeh2022} gives $N_\nu = 2.898 \pm 0.141$, hence $N_\nu < 3.180$ at two standard deviations, and converts it into a bound on any additional relativistic energy density. A gravitational wave background, whose energy density per logarithmic frequency interval in units of the critical density we write as $\Omega_{\mathrm{gw}}$, contributes as $\Omega_{\mathrm{gw}}h^2 \le (7/8)(4/11)^{4/3}\Omega_\gamma h^2\,\Delta N_\nu = 5.62\times 10^{-6}\,\Delta N_\nu$, with $\Omega_\gamma h^2 = 2.473\times 10^{-5}$~\cite{Yeh2022}. Since such a background can only add radiation, we use the weaker limit that Ref.~\cite{Yeh2022} obtains by normalizing the likelihood over $N_\nu \ge 3$, $\Delta N_\nu = N_\nu - 3 < 0.226$, which is the one it applies to a gravitational wave background; this gives $\Omega_{\mathrm{gw}}h^2 < 1.27\times 10^{-6}$, or $\Omega_{\mathrm{gw}} < 2.8\times 10^{-6}$ for $h = 0.674$. The two-sided limit, $\Delta N_\nu < 0.180$ with respect to $N_\nu = 3$, would give $1.01\times 10^{-6}$, tighter because the central value lies below 3. Converting through $S_h(f) = 3H_0^2\Omega_{\mathrm{gw}}/(2\pi^2f^3)$, the corresponding strain noise density at 999~MHz is $4.5\times 10^{-35}\,\mathrm{Hz}^{-1/2}$. One caveat applies. The bound constrains the energy density integrated over all frequencies above $\sim 2\times 10^{-11}$~Hz, the lower limit quoted in Ref.~\cite{Yeh2022}, so reading it as a bound on $\Omega_{\mathrm{gw}}(f)$ at 1~GHz assumes, as is conventional in this literature, a background occupying of order one decade there.

Against this target the array configuration falls short by a factor $1.1\times 10^{13}$ in strain as an instantaneous noise density. One year of cross-correlated observation over the 10~Hz bandwidth reduces the shortfall by the factor of Eq.~(\ref{eq:integration}) to $7.9\times 10^{10}$, and applying in addition the ceiling $\mathcal{E}_{\mathrm{max}} = 7.8$, an upper bound rather than an achievable gain, brings it to $1.0\times 10^{10}$, or $1.0\times 10^{20}$ in power. This last figure, which grants the ceiling in full, is the one by which to judge the concept, and it depends little on the choice of bound: the two-sided limit above would move it to $1.1\times 10^{10}$, and a bound of $\Omega_{\mathrm{gw}} h^2 < 1.5\times 10^{-6}$ or of $\Omega_{\mathrm{gw}} < 10^{-5}$ to $9\times 10^{9}$ or $5\times 10^{9}$ respectively.

The figure is not, however, insensitive to the choice of operating frequency. The bound falls as $f^{-3/2}$, the sensitivity rises as $f$ below the frontier frequency and as $f^{3/2}$ above it (Sec.~\ref{sec:knee}), and the integration bandwidth contributes $f^{1/4}$ the other way, so the shortfall grows as
\begin{equation}
\text{shortfall} \;\propto\; f^{9/4} \quad (T \gg T_q), \qquad
\text{shortfall} \;\propto\; f^{11/4} \quad (T \ll T_q),
\label{eq:gap_scaling}
\end{equation}
with logarithmic slopes, evaluated numerically, of 2.25 at 28~MHz and 2.75 at 10~GHz, as the exponents require. Evaluated for the same array and integration time, the shortfall runs from $1.6\times 10^{5}$ at the 3.15~MHz fundamental to $4.4\times 10^{13}$ at 10~GHz, through the $7.9\times 10^{10}$ found above at 999~MHz: a resonator of this design comes closest to the cosmological bound at the bottom of its comb, not at the top. This is no argument for a 3~MHz detector, since the frequency is set by the source one hopes to observe; Eq.~(\ref{eq:gap_scaling}) gives the price of that choice, not a case against it.

One classical parameter, the quality factor, might be expected to be limited by physics rather than by fabrication. At room temperature phonon--phonon loss in acoustic resonators is described by the Akhiezer regime; at low temperature the lifetime of thermal phonons grows and the loss passes to the Landau--Rumer description, which quartz has been observed to follow at 3.8~K, where the quality factor of these cavities is independent of frequency up to 100--200~MHz~\cite{Goryachev2013}. In that regime, $\omega\tau \gg 1$, the quality factor of an acoustic wave no longer depends on frequency, the behavior observed in these cavities, whose quality factor even increases with overtone number over a range of frequencies, which Ref.~\cite{Galliou2013} explains by phonon trapping improving with overtone number: intrinsic phonon--phonon loss does not penalize high overtones. What limits the observed $Q\times f$ at the highest overtones is Rayleigh-type scattering from dilute lattice impurities, through which the crystal behaves as a glass at high frequencies~\cite{Goryachev2013}. Whether that ceiling is irreducible is a question of crystal growth, not phonon physics, and the literature is divided: Ref.~\cite{Goryachev2013} attributes it to unavoidable crystal disorder, while a study of the same mechanism in lithium niobate calls it a non-fundamental engineering barrier, models the loss a pure crystal would show, and makes the requirement concrete: impurities of order one part per million from standard growth, against the parts-per-billion purity mitigation would demand~\cite{Campbell2025}. For the purified quartz of these devices, Ref.~\cite{Goryachev2013} puts the impurity concentration at up to 1--3 parts per billion, so the lithium niobate requirement does not transfer directly. Both may be right, the first for crystals available now, the second in principle; either way the parameters standing between this device and the bound are set by fabrication, not by the coupling. A related caution comes from the first cryogenic characterization of the MAGO prototype's mechanical eigenmodes, which found quality factors significantly below those commonly assumed in projections for that concept~\cite{MAGO2026}.

The residual factor $1.0\times 10^{10}$, obtained with the ceiling granted in full, lies entirely in the classical parameters, where, from Eq.~(\ref{eq:Sh}) and the array scaling of Sec.~\ref{sec:proposal}, $S_h \propto (m_\lambda\,\xi^2\,Q\,N)^{-1}$. Mode mass, coupling length and number of devices leave the integration bandwidth unchanged, and closing the gap through them requires a factor $1.0\times 10^{20}$ in $m_\lambda\xi^2N$: $1.0\times 10^{24}$ devices, for instance, or a coupling length grown by $1.0\times 10^{10}$ to some 80~m. Neither is an available lever, and the figures only measure the scale of the gap: at 999~MHz the coupling length of a thickness mode is bounded by the acoustic wavelength $\lambda = 6.3~\mu$m, $\ell_n \le 2\lambda/\pi^2 \approx 1.3~\mu$m (twice that when trapped), 80~m is far beyond the 30~cm gravitational wavelength within which the tidal description of Sec.~\ref{sec:BAW} holds, and $10^{24}$ devices would fill a region some 1700~km across, over which their overlap vanishes. The quality factor cannot close it, since the bandwidth $f/Q$ narrows as $Q$ grows and the integrated figure improves only as $Q^{-1/4}$ (Sec.~\ref{sec:proposal}). Nor does cooling, although with the ceiling granted it would appear to: the residual is then the thermal term alone, $2\bar{n}_{\mathrm{th}}S_h^{\mathrm{zpf}}$, and lowering the temperature would close the gap near 0.9~mK, but only through the ceiling, which Sec.~\ref{sec:ceiling} shows cannot be reached for intrinsic damping. Through the addressable fraction a readout at the standard quantum limit recovers at most its own zero-point noise, returning the one-year shortfall at best to the $7.9\times 10^{10}$ of the mode alone, which cooling below 10~mK changes by less than one percent, since the zero-point noise of the bath remains. The conclusion is therefore not that quantum technique is unhelpful. It acts on a part of the budget that is not, at present, the limiting one, and, for a resonator whose damping is intrinsic, on a smaller part than $\beta$ alone would suggest, namely the addressable fraction $\beta_a$ of Sec.~\ref{sec:ceiling}. A related conclusion follows from the transfer-function analysis of Ref.~\cite{DAgnolo2025}: on a mechanical resonance the limiting noise is typically the displacement noise of the mechanical resonator, and once noise is removed the sensitivity improves only through the stored energy and the integration time. The conclusion is also consistent with the sensitivities achieved by cavity experiments: Ref.~\cite{Berlin2022} finds that reanalysis of existing axion-haloscope data reaches dimensionless strains of order $10^{-22}$--$10^{-21}$, a projection since carried out in practice on a 2~MHz span centered at 5.311~GHz of CAPP-12T MC data, yielding 90\% exclusion limits down to $h_0 \approx 4\times 10^{-21}$~\cite{CAPP12T2026}, and Ref.~\cite{DomckeEllisRodd2025} projects $\sim 10^{-20}\,\mathrm{Hz}^{-1/2}$ broadband from a few kHz to about 10~MHz for the ADMX-EFR magnet, with $\sim 10^{-22}\,\mathrm{Hz}^{-1/2}$ at a mechanical resonance near 1~kHz. The latter is a mechanical coupling read out magnetically rather than a conversion in the cavity field, and belongs with Mechanism~A.

On the warm side of the frontier the same accounting applies with a smaller quantum share still, and the $\beta$ diagnostic points the most productive near-term investment at cryogenics, quality factor and mode mass; quantum technique becomes relevant only as cooling, or a higher operating frequency, brings $\beta$ toward unity.

\begin{figure*}
\includegraphics[width=0.8\textwidth]{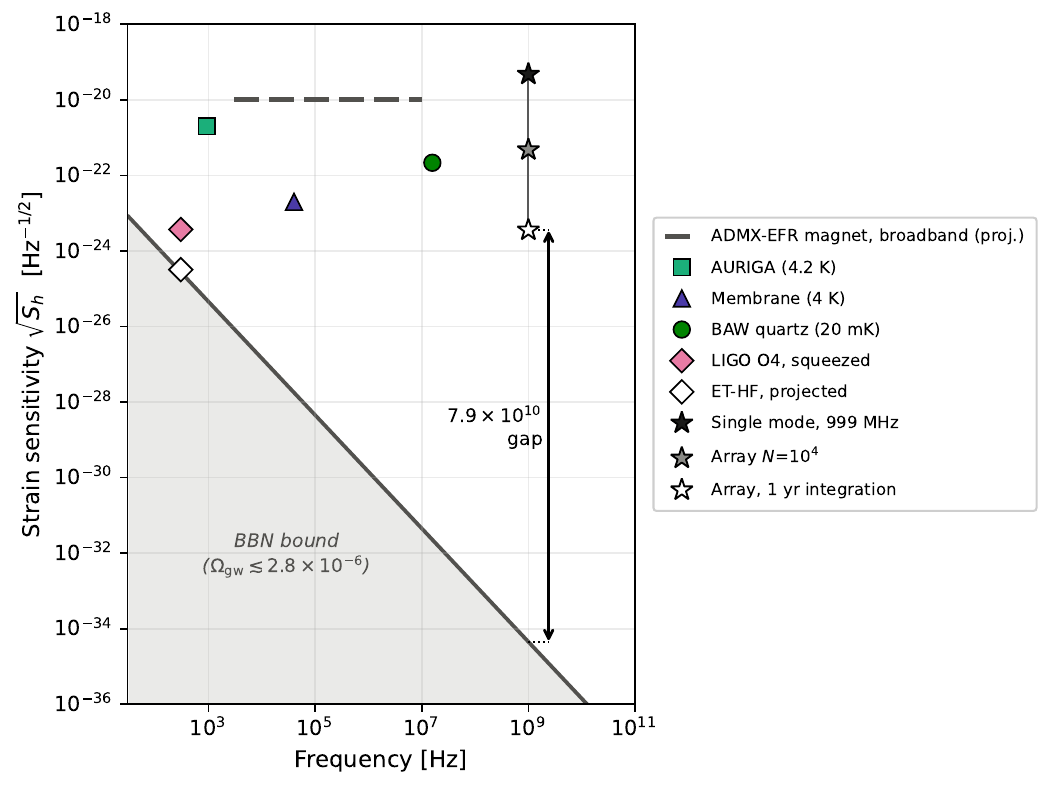}
\caption{\label{fig:landscape}Strain sensitivity of high-frequency gravitational wave detectors. Stars mark the case study of Sec.~\ref{sec:proposal}: the 999~MHz mode of a single device, as an array of $10^4$ elements, and as that array cross-correlated for one year over its 10~Hz bandwidth. The circle is the 15.8~MHz overtone of the same crystal, taken from Table~\ref{tab:landscape}. The dashed horizontal line is the broadband projection of Ref.~\cite{DomckeEllisRodd2025} for the ADMX-EFR magnet read out as a resonant mass; it is dashed because it is a projection rather than an achieved sensitivity. Of the other reference entries, AURIGA and LIGO are achieved sensitivities, while the membrane, the 15.8~MHz overtone and ET-HF are projections, as are the stars; all reference points are instantaneous noise densities. The ADMX cavity entry of Table~\ref{tab:landscape} is not shown, since Ref.~\cite{Berlin2022} reports a dimensionless strain rather than an amplitude spectral density. The shaded region below the solid line is where a cosmological background may lie given the BBN constraint $\Omega_{\mathrm{gw}} \lesssim 2.8\times 10^{-6}$; the arrow marks the shortfall of $7.9\times 10^{10}$ between the one-year array configuration and that bound. The two array stars are the best cases of Sec.~\ref{sec:proposal}, for a coherent and compact array. The three stars assume a noiseless readout: a readout adding $\nu$ zero-point units raises the single-device and array stars by $\sqrt{1+\nu/(2\bar{n}_{\mathrm{th}}+1)}$, and the one-year star, and the shortfall with it, by 1.26 at the standard quantum limit (Sec.~\ref{sec:proposal}).}
\end{figure*}

Figure~\ref{fig:landscape} places the configurations of Sec.~\ref{sec:proposal} in the context of the wider landscape. The GHz mode of the case study is the only resonant detector in the figure in the quantum-dominated regime (the interferometers reach $\beta > 1/2$ through a readout that is optical and carries negligible thermal occupation), yet it is less sensitive than the MHz overtone of the same crystal by about two orders of magnitude (a factor 220 with the quality factors and temperatures of Table~\ref{tab:landscape}, about 100 with the same quality factor and temperature), and only the integrated array configuration is competitive with the reference detectors, whose points are instantaneous noise densities. This is the cost of crossing the thermal frontier found in Sec.~\ref{sec:knee}, and the reason we present the GHz mode as a case study rather than as an optimum.

\section{Discussion}
\label{sec:discussion}

Table~\ref{tab:landscape} collects the quantum noise fraction across the landscape. Among detectors coupling tidally to a mechanical mode, $\beta$ exceeds one half only at GHz frequencies with dilution cooling. Mechanism~D cavities obey the same frontier for the same reason, their signal residing in a microwave mode with the same Bose--Einstein occupation; the ADMX entry, at 900~MHz and a cavity temperature of 150~mK, sits on the warm side of the frontier at $\beta = 0.14$. For a haloscope the amplifier chain is no small correction and Eq.~(\ref{eq:beta_readout}) is the appropriate form. The entry is thermally dominated in either case, and at the system noise reported for that run the chain reduces its $\beta$ by half, to 0.072. Interferometers are the exception, holding $\beta \sim 0.85$--0.95 in their shot-noise-dominated band: an optical mode carries negligible thermal occupation, and the quantum share of their budget is set by shot noise.

The asymmetry has one origin, the fluctuation--dissipation theorem: a mechanical coupling that grows with the mass carries a thermal force noise that grows with it and is proportional to $k_BT$, so the only route to $\beta > 1/2$ for the mode of a resonant mass is $\hbar\omega > k_BT\ln 3$, and Sec.~\ref{sec:knee} shows what reaching that regime costs. Light propagation is exempt because optical photons carry negligible thermal occupation; that is the structural reason interferometers have benefited from squeezing while resonant masses have not. For a resonant mass the same theorem also fixes what crossing the frontier buys: with intrinsic damping the bath's zero-point force noise survives every operation on the mode, and the share a quantum technique can remove is not $\beta$ but the addressable fraction $\beta_a$ of Eq.~(\ref{eq:beta_addressable}): for the case study at the standard quantum limit, an available improvement of 1.41 against a ceiling of 11.0.

\emph{Establishing a signal.} The array is not only a sensitivity multiplier. In a single device a stationary background is indistinguishable from the device's own noise floor unless the floor is known absolutely to the same precision, which these levels exclude. The signal must then be recovered from the correlation between detectors with independent noise~\cite{AllenRomano1999}. The nanohertz band supplies the precedent: a common spectrum across a pulsar array does not by itself establish a gravitational-wave origin, and the evidence for one rests on the Hellings--Downs correlation~\cite{HellingsDowns1983}, a universal, quasi-quadrupolar function of the angular separation between pulsars~\cite{NANOGrav15yr}. The present framework adds an internal test of the same kind, transposed from sky angle to overtone number: a broadband signal must drive every odd overtone with on-resonance amplitude $x_n = \tfrac{1}{2}h\,\ell_n Q_n$ (for a given spectral shape, a relative response fixed by $\ell_n \propto n^{-2}$ and the measured $Q_n$ once the plate thickness is known), and an environmental disturbance has no reason to follow it. We record this as a requirement on any candidate excess rather than as a strategy: the gap is far too large for the question to be practical here.

A full budget would carry further noise sources. Losses in the transduction and amplification chain contribute to the excess readout noise, covered by the $\nu_{\mathrm{exc}}$ of Eq.~(\ref{eq:beta_readout}); transients such as those observed in a BAW antenna, of undetermined origin, appear as non-Gaussian outliers and are separable from a stationary background, and muon coincidence detectors have been proposed to identify cosmic-ray events~\cite{Goryachev2021}; cryostat vibrations couple parametrically, through acoustoelastic modulation of the elastic constants by quasi-static strain. The last requires the fractional frequency modulation to stay within the fractional linewidth $1/Q = 10^{-8}$: with an acoustoelastic coefficient of order unity, the quasi-static strain in the crystal must remain below $\sim 10^{-8}$ over the measurement time. This is a specification for the mount and the vibration isolation, not a property we have established. None alters the conclusions; the readout term is the one that matters, and Eq.~(\ref{eq:joint}) is the form in which to specify it.

Two directions follow. The first is the measurement of Sec.~\ref{sec:knee}, within reach of existing apparatus: the invariant of Eq.~(\ref{eq:invariant}), the increase across the frontier of the logarithmic slope of $\sqrt{S_hQ_n}$, one half for the noise of the mode between sampling points far from it, obtained from the noise floors and linewidths of odd overtones spanning the frontier on one crystal. Dividing out the measured $Q_n$ makes the result independent of the frequency dependence of the quality factor, which displaces the absolute exponents and, if it changes between the sampling points, the slope difference of $\sqrt{S_h}$ alone. Repeating at a second temperature moves the frontier by a known amount and checks that the slope is a universal function of $T/T_q$: it would test the Bose--Einstein scaling on which the frontier is placed, which is a property of the mode and not a convention. The second direction concerns the readout: the requirement of Eq.~(\ref{eq:joint}) suggests that effort directed at the excess noise of the readout chain is, for a mode already in its ground state, worth more than further cooling, since with the chains demonstrated so far it is the largest term of the budget and one that can still be reduced.

\section*{Data availability statement}

The data that support the findings of this study are openly available in Zenodo at \href{https://doi.org/10.5281/zenodo.19862916}{doi:10.5281/zenodo.19862916}, together with the Python scripts that generate the figures and reproduce every numerical result.

\begin{acknowledgments}
The author acknowledges the use of Anthropic Claude as a writing assistant. The author checked this material and takes full responsibility for the content of the article. This research received no external funding.
\end{acknowledgments}

\end{document}